# Technical Report – CVE-2022-46480, CVE-2023-26941, CVE-2023-26942, and CVE-2023-26943

## Introduction

The following technical report provides background information relating to four CVEs found in the following products:

Ultraloq UL3 BT (2nd Generation) (Firmware 02.27.0012) – CVE-2022-46480

Yale Conexis L1 Smart Lock (Firmware v1.1.0) – CVE-2023-26941

Yale IA-210 Intruder Alarm (Firmware v1.0) – CVE-2023-26942

Yale Keyless Smart Lock (Firmware v1.0) – CVE-2023-26943

The work discussed here was carried out by Ash Allen, Dr. Alexios Mylonas, and Dr. Stilianos Vidalis as part of a wider research project into smart device security. Responsible disclosure of all four issues has been made with the appropriate vendors, and they have been acknowledged as vulnerabilities.

## CVE-2022-46480

The Ultraloq UL3 BT (2nd Generation) (hereafter referred to as the UL3) is one of the most popular smart locks in the United States ("#1 Selling Smart Lock Online" - https://u-tec.com/). Previous non-academic research[1] discusses flaws found with the companion app, and the ability to brute-force the Bluetooth encryption. Our work uncovered an even more fundamental flaw with the Bluetooth implementation. The device unlocks by receiving a 16-byte value to a service specified via the GATT (General ATTribute Profile). Our work uncovers that this value is changed on a per session basis, rather than per unlock event. This means that whilst a session remains open, unlock commands using replayed data will be successful, even if they do not come from the original client.

**2022.11.15 17:59:32.844 | < C | 7200 | 7201 | 08c71149fb1a7105298eaf175bf5166b ( I q ) [ k)**
2022.11.15 17:59:33.047 | > N | 7200 | 7201 | 5970e43108c32af5811fbfb3bfce5400 (Yp 1 * T )
2022.11.15 17:59:33.766 | > N | 7200 | 7201 | f086584891767d5f32b22674bb2dcb49 ( XH v}_2 &t - I)
2022.11.15 17:59:38.762 | > N | 7200 | 7201 | 2b2dcdae303f8df4fb109e818189fb74 (+- 0? t)
**2022.11.15 17:59:41.244 | < C | 7200 | 7201 | 08c71149fb1a7105298eaf175bf5166b ( I q ) [ k)**
2022.11.15 17:59:41.446 | > N | 7200 | 7201 | 5970e43108c32af5811fbfb3bfce5400 (Yp 1 * T )
2022.11.15 17:59:42.167 | > N | 7200 | 7201 | f086584891767d5f32b22674bb2dcb49 ( XH v}_2 &t - I)
2022.11.15 17:59:47.147 | > N | 7200 | 7201 | 2b2dcdae303f8df4fb109e818189fb74 (+- 0? t)
**2022.11.15 17:59:49.644 | < C | 7200 | 7201 | 08c71149fb1a7105298eaf175bf5166b ( I q ) [ k)**
2022.11.15 17:59:49.847 | > N | 7200 | 7201 | 5970e43108c32af5811fbfb3bfce5400 (Yp 1 * T )
2022.11.15 17:59:50.567 | > N | 7200 | 7201 | f086584891767d5f32b22674bb2dcb49 ( XH v}_2 &t - I)
2022.11.15 17:59:55.547 | > N | 7200 | 7201 | 2b2dcdae303f8df4fb109e818189fb74 (+- 0? t)
**2022.11.15 17:59:57.263 | < C | 7200 | 7201 | 08c71149fb1a7105298eaf175bf5166b ( I q ) [ k)**
2022.11.15 17:59:57.467 | > N | 7200 | 7201 | 5970e43108c32af5811fbfb3bfce5400 (Yp 1 * T )
2022.11.15 17:59:58.187 | > N | 7200 | 7201 | f086584891767d5f32b22674bb2dcb49 ( XH v}_2 &t - I)
2022.11.15 18:00:03.168 | > N | 7200 | 7201 | 2b2dcdae303f8df4fb109e818189fb74 (+- 0? t)

---

[1] Spring, T., 2019. *Smart Lock Turns Out to be Not So Smart, or Secure* - https://threatpost.com/smart-lock-turns-out-to-be-not-so-smart-or-secure/146091/

Furthermore, our work uncovered poor session management while a user attempts to unlock the device. We were able to force sessions to remain open for more than 12 minutes, giving an attacker plenty of opportunity to exploit the vulnerability.

2022.11.19 **15:04:47.385** | < C | 7200 | 7201 | **8e91fb970f0bbdac85906943b9c60a90 ( iC )**
2022.11.19 **15:17:05.384** | < C | 7200 | 7201 | **8e91fb970f0bbdac85906943b9c60a90 ( iC )**

## CVE-2023-26941

The Yale Conexis L1 was until recently the flagship model of the Yale smart lock range (the Conexis L2 has now taken that spot). Winning plaudits in many consumer surveys[234], the Conexis L1 has a significant number of installed units. The Conexis L1 uses standard Mifare Classic 1K cards (larger cards can be provisioned, but the extra space is not required and is not used). Initial setup is done using a master key card. We were able to crack the encryption on the card using a Proxmark3. Easy RFID development tool and dump the contents to a text file. The first seven blocks on the card use non-standard encryption keys (the default is FFFFFFFFFFFF). We were able to recover these within 20 seconds in almost every case. Data was written in several places, all of which protected by the non-default keys. During the initial pairing process a counter is incremented in block 2. This is incremented again after each factory reset of the device, as demonstrated below:

```
"blocks": {
"0": "95E43AA5EE08040002E9981FA7F5D11D",
"1": "095FCA99D806ECCEB9328A6466CA3D10",
"2": "080807000000000000000000000000",

"blocks": {
"0": "95E43AA5EE08040002E9981FA7F5D11D",
"1": "095FCA99D806ECCEB9328A6466CA3D10",
"2": "090907000000000000000000000000",
```

Of greater interest is a batch of data written to a pair of neighboring blocks, the locations of which are chosen at random during the initial pairing of the RFID tag. These values are updated in a "tick-tock" manner during the unlock process:

```
"4": "F55129991B000000000000000000000",
"5": "60E0E0E087000000000000000000000",

"4": "FF931B6B23000000000000000000000",
"5": "60E0E0E087000000000000000000000",

"4": "FF931B6B23000000000000000000000",
"5": "E6EEBEDE6F000000000000000000000",

"4": "3070F0F047000000000000000000000",
"5": "E6EEBEDE6F000000000000000000000",
```

If the values in these blocks are altered in any way, then the card will not work. We were able to create duplicate cards from the text file dump and were able to successfully open the lock. Additionally, as data on the duplicate card was updated as part of the unlock process, the original card is locked out and no longer works. The reverse is also true – if the original card is used before the duplicate then the duplicate will fail to open the lock.

---

[2] https://diyworks.co.uk/best-smart-lock/
[3] https://www.goodhousekeeping.com/uk/product-reviews/tech/g40557412/best-smart-locks/
[4] https://www.t3.com/features/best-smart-lock

## CVE-2023-26942

The Yale IA-210 intruder alarm is a consumer-grade device designed for home installation. The keypad supports both PIN entry and RFID tag disarming. Investigating the tag provided with the device, we attempted to dump the contents to a text file. Investigation showed that all sectors were using the default encryption key (the sector 0 key can be seen in block 3):

```
[=] ----+-------------------------------------------+
[=] blk | data                                      |
[=] ----+-------------------------------------------+
[=]   0 | 3D 06 CD 45 B3 88 04 00 C8 42 00 20 00 00 00 16 |
[=]   1 | 00 00 00 00 00 00 00 00 00 00 00 00 00 00 00 00 |
[=]   2 | 00 00 00 00 00 00 00 00 00 00 00 00 00 00 00 00 |
[=]   3 | FF FF FF FF FF FF FF 07 80 69 FF FF FF FF FF FF |
```

This makes cloning the fob incredibly easy – it is possible to clone the tag with an NFC-enabled mobile phone or a dedicated cloning tool with just a couple of seconds access to the tag. The only authentication step performed by the alarm is to ensure that the tag can be decrypted using the default keys, and that the contents of block 0 also match. This block is read-only and contains data that can be used to identify the tag. Special tags known as "magic" tags are available, and these allow the rewriting of this sector. This is how we are able to duplicate the original tag.

## CVE-2023-26943

The Yale Keyless Smart Lock is a mid-range device that supports PIN entry, RFID, and 868MHz RF remote control via an optional add-in module. As with the Conexis L1, this lock uses RFID tags with a non-default set of encryption keys for the first seven sectors, as shown below:

```
[+] -----+-----+--------------+---+--------------+----
[+]  Sec | Blk | key A        |res| key B        |res
[+] -----+-----+--------------+---+--------------+----
[+]  000 | 003 | 681E9E9B3FE9 | N | FFFFFFFFFFFF | D
[+]  001 | 007 | ADAE73113441 | N | FFFFFFFFFFFF | D
[+]  002 | 011 | 6C6FAAC8E598 | N | FFFFFFFFFFFF | D
[+]  003 | 015 | BFBCF91B36CB | N | FFFFFFFFFFFF | D
[+]  004 | 019 | 58599AF4D3A4 | N | FFFFFFFFFFFF | D
[+]  005 | 023 | 828340E60956 | N | FFFFFFFFFFFF | D
[+]  006 | 027 | 5F5E9D3BD48B | N | FFFFFFFFFFFF | D
[+]  007 | 031 | FFFFFFFFFFFF | D | FFFFFFFFFFFF | D
[+]  008 | 035 | FFFFFFFFFFFF | D | FFFFFFFFFFFF | D
[+]  009 | 039 | FFFFFFFFFFFF | D | FFFFFFFFFFFF | D
[+]  010 | 043 | FFFFFFFFFFFF | D | FFFFFFFFFFFF | D
[+]  011 | 047 | FFFFFFFFFFFF | D | FFFFFFFFFFFF | D
[+]  012 | 051 | FFFFFFFFFFFF | D | FFFFFFFFFFFF | D
[+]  013 | 055 | FFFFFFFFFFFF | D | FFFFFFFFFFFF | D
[+]  014 | 059 | FFFFFFFFFFFF | D | FFFFFFFFFFFF | D
[+]  015 | 063 | FFFFFFFFFFFF | D | FFFFFFFFFFFF | D
[+] -----+-----+--------------+---+--------------+----
```

Unlike the Conexis L1, however, no extra data is written on the tag. During authentication the tag is decrypted using the given encryption keys, then checking that the data in block 0 matches that stored during the initial pairing of the tag. We were able to consistently duplicate these tags with approximately 20 seconds of access to the original tag. Furthermore, once duplicated, no indication of duplication is given to the original user. There is no audit log available, and users are not alerted of duplicate tag usage.

Unlike with the Conexis L1, there is no tripwire (the "tick-tock" updates) that can be triggered to indicate that a duplicate card exists.